\documentclass[a4paper]{jpconf}
\usepackage{graphicx}
\newcommand{\1}{\vec}
\newcommand{\2}{\partial}

\newcommand{\ter}{Terzan~5~}
\newcommand{\beq}{\begin{equation}}
\newcommand{\eeq}{\end{equation}}

\begin{document}
\title{Modelling the average spectrum expected from a population of gamma-ray globular clusters}

\author{C Venter$^1$ and A Kopp$^2$}

\address{$^1$Centre for Space Research, North-West University, Potchefstroom Campus, Private Bag X6001, Potchefstroom 2520, South Africa\\
$^2$Institut f\"{u}r Experimentelle und Angewandte Physik, Christian-Albrechts-Universit\"{a}t zu Kiel, Leibnizstrasse 11, 24118 Kiel, Germany}

\begin{abstract}
Millisecond pulsars occur abundantly in globular clusters. They are expected to be responsible for several spectral components in the radio through $\gamma$-ray waveband (e.g., involving synchrotron and inverse Compton emission), as have been seen by Radio Telescope Effelsberg, \textit{Chandra} X-ray Observatory, \textit{Fermi} Large Area Telescope, and the High Energy Stereoscopic System (H.E.S.S.) in the case of Terzan~5 (with fewer spectral components seen for other globular clusters). H.E.S.S.\ has recently performed a stacking analysis involving 15 non-detected globular clusters and obtained quite constraining average flux upper limits above 230~GeV. We present a model that assumes millisecond pulsars as sources of relativistic particles and predicts multi-wavelength emission from globular clusters. We apply this model to the population of clusters mentioned above to predict the average spectrum and compare this to the H.E.S.S.\ upper limits. Such comparison allows us to test whether the model is viable, leading to possible constraints on various average cluster parameters within this framework.
\end{abstract}

\section{Introduction}
There are nearly 160 Galactic globular clusters (GCs) known~\cite{Harris96}. Each consists of hundreds of thousands of stars held together by their mutual gravity, and they orbit the Galactic Centre in a spherical distribution. GCs are typically about ten gigayears old and are therefore expected to harbour many evolved stellar objects, since the latter should have had ample time to complete their evolutionary processes. The high stellar densities in the cores of GCs also enhance stellar encounter rates, facilitating the formation of objects such as low-mass X-ray binaries (LMXRBs), cataclysmic variables, white dwarfs, and pulsars~\cite{Pooley03}. LMXRBs are believed to be the progenitors of millisecond pulsars (MSPs; \cite{Alpar82}), and since they occur abundantly in GCs, the same should hold true for MSPs. This is indeed found to be the case: 28 of the Galactic GCs contain more than 144 confirmed radio pulsars\footnote{http://www.naic.edu/$\sim$pfreire/GCpsr.html}, the bulk of these being MSPs. It was furthermore estimated that there should be $2~600-4~700$ Galactic GC MSPs observable in $\gamma$-rays~\cite{Abdo10}. 

GCs are multi-wavelength objects, being visible from radio to the highest energies. For example, several radio structures are visible in the direction of \ter at 11~cm and 21~cm~\cite{Clapson11}, while diffuse X-ray emission has also been detected from this GC~\cite{Clapson11,Eger10}. The \textit{Fermi} Large Area Telescope (LAT) plausibly detected about a dozen GCs in the GeV energy band~\cite{2FGL}, and their spectral characteristics point to the cumulative emission from a population of GC MSPs. In the TeV domain, H.E.S.S.\ has published upper limits for 47~Tucanae~\cite{Aharonian09_Tuc}, and detected a very-high-energy (VHE) excess in the direction of \ter\cite{Abramowski11}.

Several models have been proposed to explain the observed GC spectra. The total GeV contribution from GC MSPs was estimated by summing up individual predicted pulsed curvature radiation (CR) spectra from a population of MSPs~\cite{HUM05,Venter08}. An alternative scenario~\cite{Cheng10} assumed that the GeV emission was due to inverse Compton (IC) radiation by leptons escaping from the MSP magnetospheres, upscattering Cosmic Microwave Background (CMB), stellar, as well as Galactic background (infrared and optical) photons. Another model~\cite{BS07} considered MSPs that accelerate leptons either at the shocks originating during collisions of neighbouring pulsar winds or inside the pulsar magnetospheres. These leptons escape from the magnetospheres and diffuse through the GC, encountering target photon fields such as optical and CMB emission. The latter are upscattered via the IC process, leading to GeV / TeV emission. This model was extended~\cite{Venter09_GC,Zajczyk13} using alternative particle injection spectra, and also calculating the expected synchrotron radiation (SR) from leptons moving in a homogeneous cluster magnetic field $B$. Recently, the model was significantly refined~\cite{Kopp13} and now includes a line-of-sight calculation of the X-ray surface brightness (which is used to constrain the diffusion coefficient) as well as full particle transport, assuming spherical symmetry and a steady-state regime. Apart from leptonic models, there two other ideas. Non-accreting white dwarfs may make a considerable contribution to the cumulative $\gamma$-ray flux seen from GCs~\cite{Bednarek12}, depending on their abundance. Lastly, a model invoking $\gamma$-ray burst remnants as sources of energetic leptons and hadrons was also put forward~\cite{Domainko11}. A short burst may accelerate hadrons, which may in turn collide with ambient target nuclei, leading to $\pi^0$ particles that eventually decay into $\gamma$-rays. In this model, X-rays may result from IC on optical stellar photons by primary electrons accelerated by the relativistic blast wave.

We note that a search for diffuse X-ray emission from several {\it Fermi}-detected GCs~\cite{Eger12} failed to detect any such emission above the Galactic background level, although a recent reanalysis of the archival \textit{Chandra} data detected a new diffuse X-ray emission feature within the half-mass radius of 47 Tucanae \cite{Wu14}. In the context of the leptonic MSP models, these upper limits and detection place constraints on parameters such as the number of embedded MSPs $N_{\rm MSP}$ as well as the typical cluster field strength $B$. One way to help discriminate between the various models would be to follow a population approach. H.E.S.S.\ has recently obtained an  upper limit to the average TeV flux of 15 non-detected GCs~\cite{Abramowski13}. They noted that their upper limit was lower than the flux predicted by a simple leptonic scaling model by a factor of $\sim3 - 30$, depending on model assumptions. This paper represents a first attempt to model the average TeV flux from the population of 15 GCs using our refined leptonic GC model. Our motivation is to compare our results with those of the scaling model, and second, to assess the plausibility of the MSPs scenario by testing whether our average spectrum satisfies the TeV upper limits. In Section~\ref{sec:model}, we briefly discuss our model, after which follow a description of the assumed parameters and calculation method (Section~\ref{sec:cum}), as well as our conclusions (Section~\ref{sec:concl}).

\section{The basic GC model}
\label{sec:model}

We solve the following transport equation numerically
\beq
\frac{\2 n_{\rm e}}{\2t}=\1\nabla\cdot\left({\cal K}\cdot\1\nabla n_{\rm e} \right)-\frac{\2}{\2E_{\rm e}}\left(\dot{E}_{\rm e} n_{\rm e}\right)+Q,\label{eq:transport}
\eeq
with $n_{\rm e}$ the electron\footnote{We use the word `electron' in a wider sense. Pair production may take place, additionally providing the possibility of positrons as efficient radiators.} density,  which is a function of the radius vector $\1r_{\rm s}$, $E_{\rm e}$ the electron energy, ${\cal K}$ the diffusion tensor, $\dot{E}_{\rm e}>0$ the particle energy losses, and $Q$ the source term. We assume a source term located at $\1r_{\rm s}=\1r_{\rm C}$ that is of the form
\beq
Q=Q_0{E_{\rm e}}^{-\Gamma}\delta(\1r_{\rm s}-\1r_{\rm C}),
\eeq
with the spectral index $\Gamma$. We assume stationarity and spherical symmetry, leading to
\beq
0=\frac{1}{{r_{\rm s}}^2}\frac{\2}{\2r_{\rm s}}\left({r_{\rm s}}^2 \kappa \frac{\2n_{\rm e}}{\2r_{\rm s}}\right)
-\frac{\2}{\2E_{\rm e}}\left(\dot{E}_{\rm e} n_{\rm e}\right)+Q,
\eeq
with $\kappa$ the scalar diffusion coefficient. This may be rewritten as 
\beq
\frac{\2 n_{\rm e}}{\2E_{\rm e}}=\frac{1}{\dot{E}_{\rm e}}\left(\frac{1}{{r_{\rm s}}^2}\frac{\2}{\2r_{\rm s}}\left({r_{\rm s}}^2 \kappa \frac{\2n_{\rm e}}{\2r_{\rm s}}\right)-n_{\rm e}\frac{\2\dot{E}_{\rm e}}{\2E_{\rm e}}+Q\right),
\eeq
and is solved numerically using a Crank-Nicolson algorithm.
Our particle injection spectrum  is normalised as follows:
\beq
\int_{E_{\rm e,min}}^{E_{\rm e,max}}E_{\rm e}Q\,dE_{\rm e} = N_{\rm MSP}\eta\langle L_{\rm sd}\rangle,\label{eq:qnorm}
\eeq
with $\eta$ the particle conversion efficiency, and $\langle L_{\rm sd}\rangle$ the average MSP spin-down luminosity. After solving for $n_{\rm e}$, we calculate the SR and IC fluxes as detailed in~\cite{Kopp13}. We assume Bohm diffusion
\beq
\kappa(r_{\rm s},E_{\rm e})=\frac{cE_{\rm e}}{3eB(r_{\rm s})},
\eeq
where $c$ and $e$ denote speed of light and electron charge. As an alternative, we also investigated a coefficient of the form
\beq
\kappa(r_{\rm s},E_{\rm e})=\kappa_0(r_{\rm s})\left(\frac{E_{\rm e}}{E_{\rm e,0}}\right)^{\alpha},\label{eq:kappa} 
\eeq
with $E_{\rm e,0}=1$ TeV and $\alpha=0.6$. For simplicity, we only considered spatially constant $\kappa_0$ and $B$.

\begin{center}
\lineup
\begin{table}[b]
\caption{Assumed GC parameters.}\label{tab1}
%\footnotesize\rm
\centering
\begin{tabular}{@{}*{8}{l}}
\br
GC Name & $d$ & $N_*$ & $N_{\rm MSP}$ & $Q_0$               & $r_{\rm c}$ & $r_{\rm hm}$ & $r_{\rm t}$\\
        & (kpc) & $(10^5)$ &          & $(10^{32}$/erg/s)   & ($^\prime$) & ($^\prime$)  & ($^\prime$)\\
\mr
NGC~104 (47~Tuc) & \04.5  & 4.57  & $33^{+15}_{-15}$    & \09.55 & 0.36  & 3.17 & 42.86 \\
NGC~6388         & \09.9  & 5.8   & $180^{+120}_{-100}$ & 52.1   & 0.12  & 0.52 & \06.21  \\
NGC~7078         & 10.4   & 4.13  & 25 $(<56)$          & \07.24 & 0.14  & 1.00 & 21.5  \\
Terzan~6         & \06.8  & 0.29  & 25                  & \07.24 & 0.05  & 0.44 & 17.39 \\
Terzan~10        & \05.8  & 0.38  & 25                  & \07.24 & 0.9   & 1.55 & \05.06  \\
NGC~6715         & 26.5   & 4.79  & 25                  & \07.24 & 0.09  & 0.82 & \07.47  \\
NGC~362          & \08.6  & 1.58  & 25                  & \07.24 & 0.18  & 0.82 & 16.11 \\
Pal~6            & \05.8  & 0.31  & 25                  & \07.24 & 0.66  & 1.2  & \08.36  \\
NGC~6256         & 10.3   & 0.21  & 25                  & \07.24 & 0.02  & 0.86 & \07.59  \\
Djorg~2          & \06.3  & 1.0   & 25                  & \07.24 & 0.33  & 1.0  & 10.53 \\
NGC~6749         & \07.9  & 0.24  & 25                  & \07.24 & 0.62  & 1.1  & \05.21  \\
NGC~6144         & \08.9  & 0.48  & 25                  & \07.24 & 0.94  & 1.63 & 33.25 \\
NGC~288          & \08.9  & 0.32  & 25                  & \07.24 & 1.35  & 2.23 & 12.94 \\
HP~1             & \08.2  & 0.48  & 25                  & \07.24 & 0.03  & 3.1  & \08.22  \\
Terzan~9         & \07.1  & 0.02  & 25                  & \07.24 & 0.03  & 0.78 & \08.22  \\
\br
\end{tabular}
\end{table}
\end{center}

\section{Estimating the average GC spectrum - a first approach}
\label{sec:cum}
The model described in Section~\ref{sec:model} has been successfully applied to the case of Terzan~5~\cite{Kopp13}, where we performed a line-of-sight integration of the X-ray flux in order to constrain the lepton diffusion coefficient. We found that values of $\kappa_0 \approx10^{-4}$~kpc$^2$Myr$^{-1}$ gave good fits. In this paper, we apply the same model to 15 non-detected GCs~\cite{Abramowski13}, using fixed parameters as noted in table~\ref{tab1}. We have used the values of~\cite{Abdo10} for $N_{\rm MSP}$ where possible, and $N_*$ values from \cite{Lang93}, and obtained distances $d$ and structural parameters\footnote{See also http://gclusters.altervista.org/index.php} $r_{\rm c}$ (core radius), $r_{\rm hm}$ (half-light radius, used as a proxy for half-mass radius), and $r_{\rm t}$ (tidal radius) from~\cite{Harris96}. Typical values of $\eta\sim0.01$, $N_{\rm MSP}\sim25$, and $\langle L_{\rm sd}\rangle\sim2\times10^{34}$~erg\,s$^{-1}$ led to values for the source strength $Q_0$. We have used $B=5~\mu$G and $\Gamma=2.0$ throughout, and used optical GC plus CMB photons as background fields for our IC calculation (assuming an average stellar temperature of $T=4~500$~K).

\begin{figure}[t]
\begin{minipage}{17pc}
\includegraphics[width=17pc]{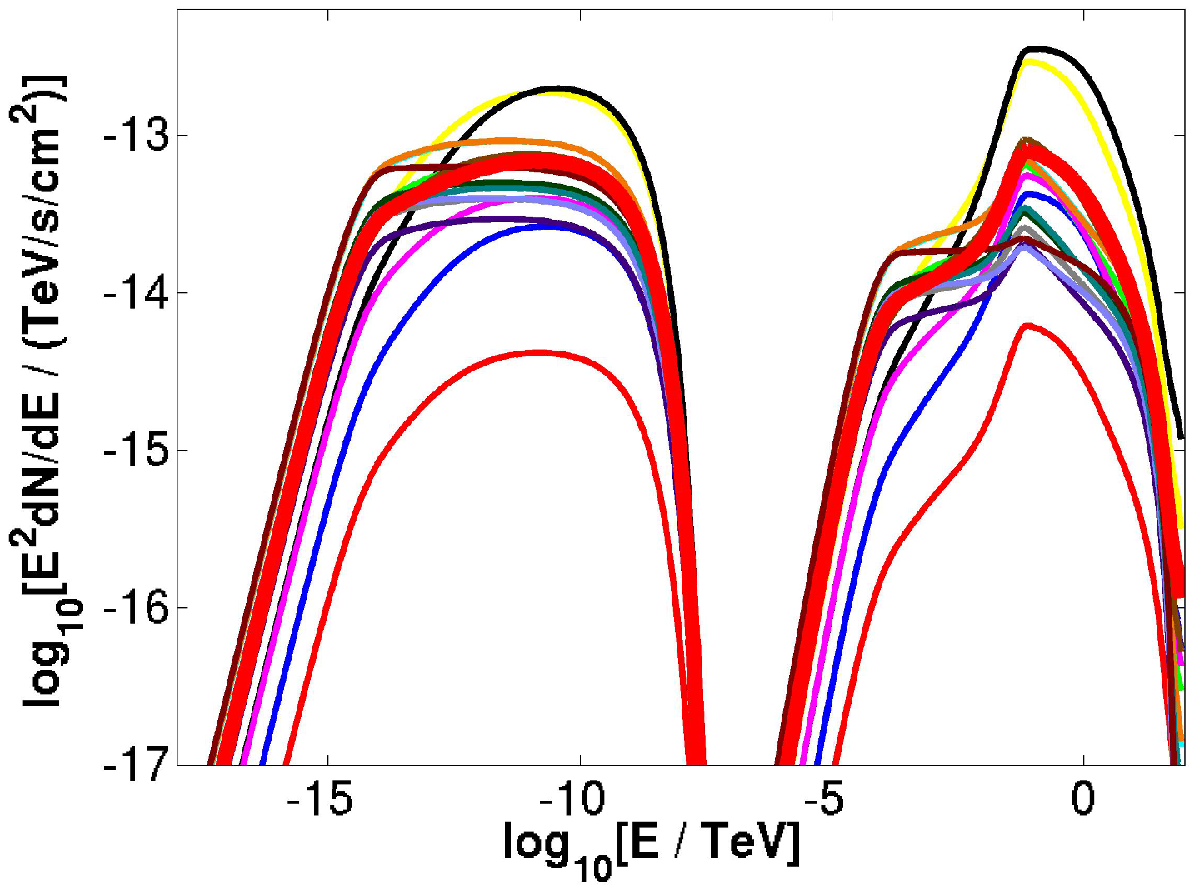}
\caption{\label{fig1} Predicted differential spectra $E^2dN/dE$ for 15 single GCs (thin lines), as well as the average spectrum (thick red line). The two components represent the SR and IC spectra.}
\end{minipage}\hspace{2pc}%
\begin{minipage}{17pc}
\includegraphics[width=17pc]{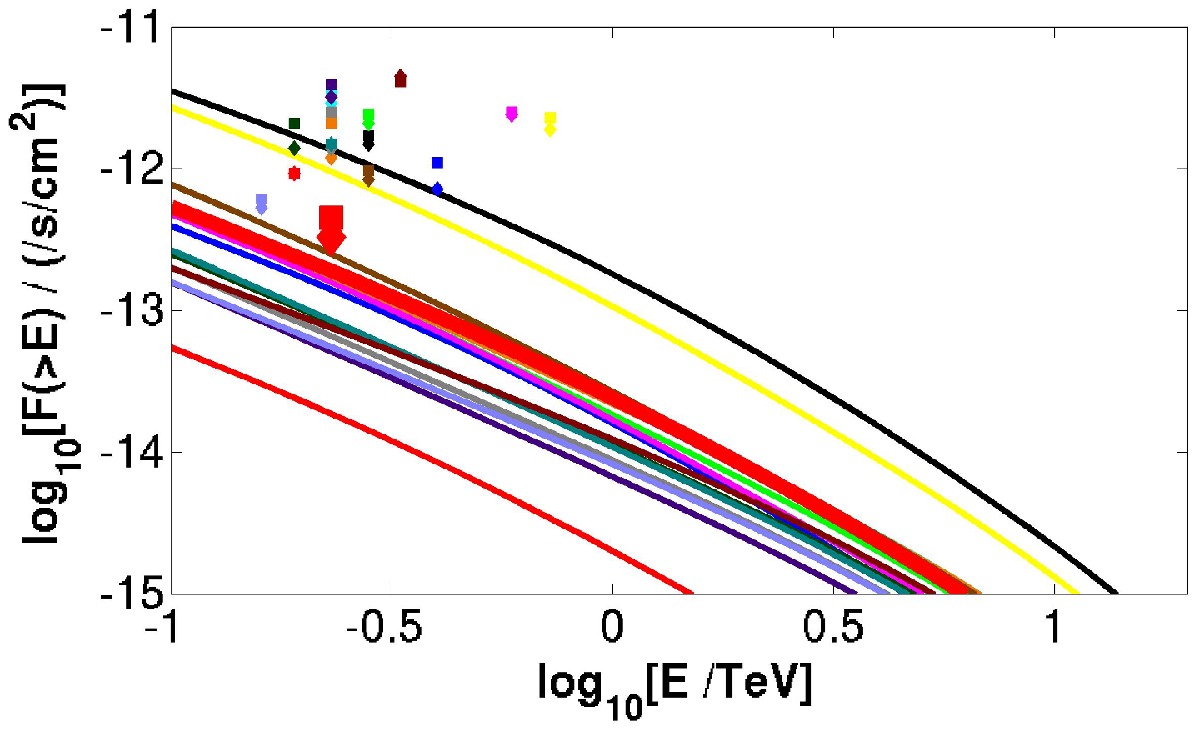}
\caption{\label{fig2}Same as figure~\ref{fig1}, but for integral flux $F(>\!\!E)$. Upper limits for the point-like source analysis are indicated by diamonds, and those for the extended source analysis by squares~\cite{Abramowski13}. Associated spectra and upper limits have the same colour. The larger (lowest) symbols are for the average spectrum.}
\end{minipage} 
\end{figure}

We first produced SR and IC spectra for each individual cluster. We next added these spectra and divided by the number of GCs to obtain the average predicted spectrum. This is indicated in figure~\ref{fig1} (differential flux) and figure~\ref{fig2} (integral flux). None of the single-cluster spectra violate the TeV upper limits. The stacked upper limits are $F(>0.23~{\rm TeV}) = 3.3\times10^{-13}$~cm$^{-2}$s$^{-1}$ for a point-like source analysis, and $F(>0.23~{\rm TeV}) = 4.5\times10^{-13}$~cm$^{-2}$s$^{-1}$ for an extended source analysis~\cite{Abramowski13}. Our average spectrum gives $F(>0.23~{\rm TeV}) \approx 2\times10^{-13}$~cm$^{-2}$s$^{-1}$, which satisfies these observational upper limits. However, we note that different choices of parameters will lead to average fluxes that could even exceed these limits in some cases.
%We note that This violation is within the range predicted by the scaled model that only takes GC starlight photons into account (a factor 18.5, which becomes 13$-$73 when errors are taken into account for the point-like analysis; a factor 13.3, or 9$-$54 with errors, for the extended source analysis).

\begin{figure}[t]
\begin{minipage}{16pc}
\includegraphics[width=16pc]{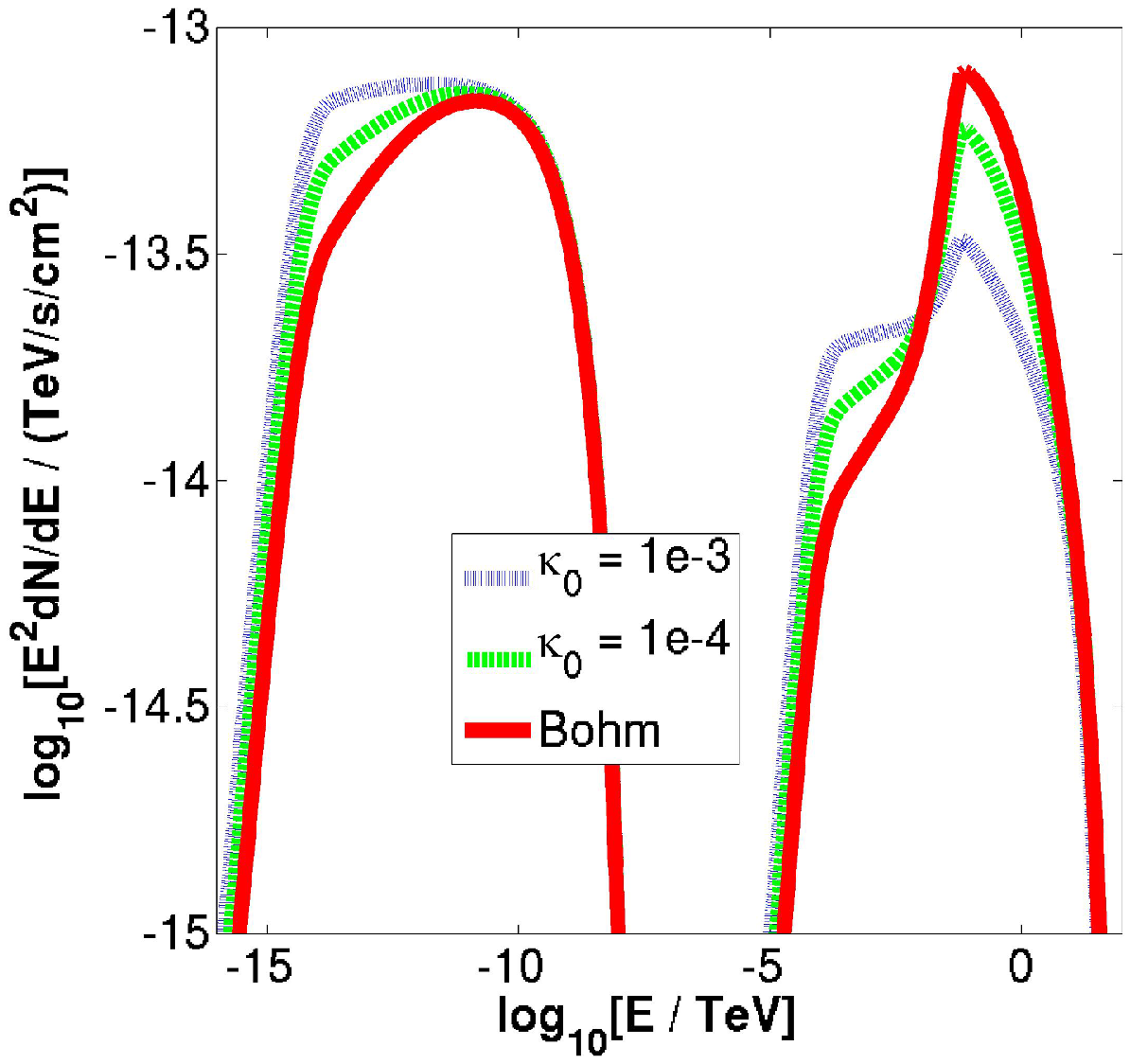}
\caption{\label{fig3}Predicted average differential spectra. The line types indicate different diffusion coefficient assumptions as noted in the legend.}
\end{minipage}\hspace{2pc}%
\begin{minipage}{16pc}
\includegraphics[width=16pc]{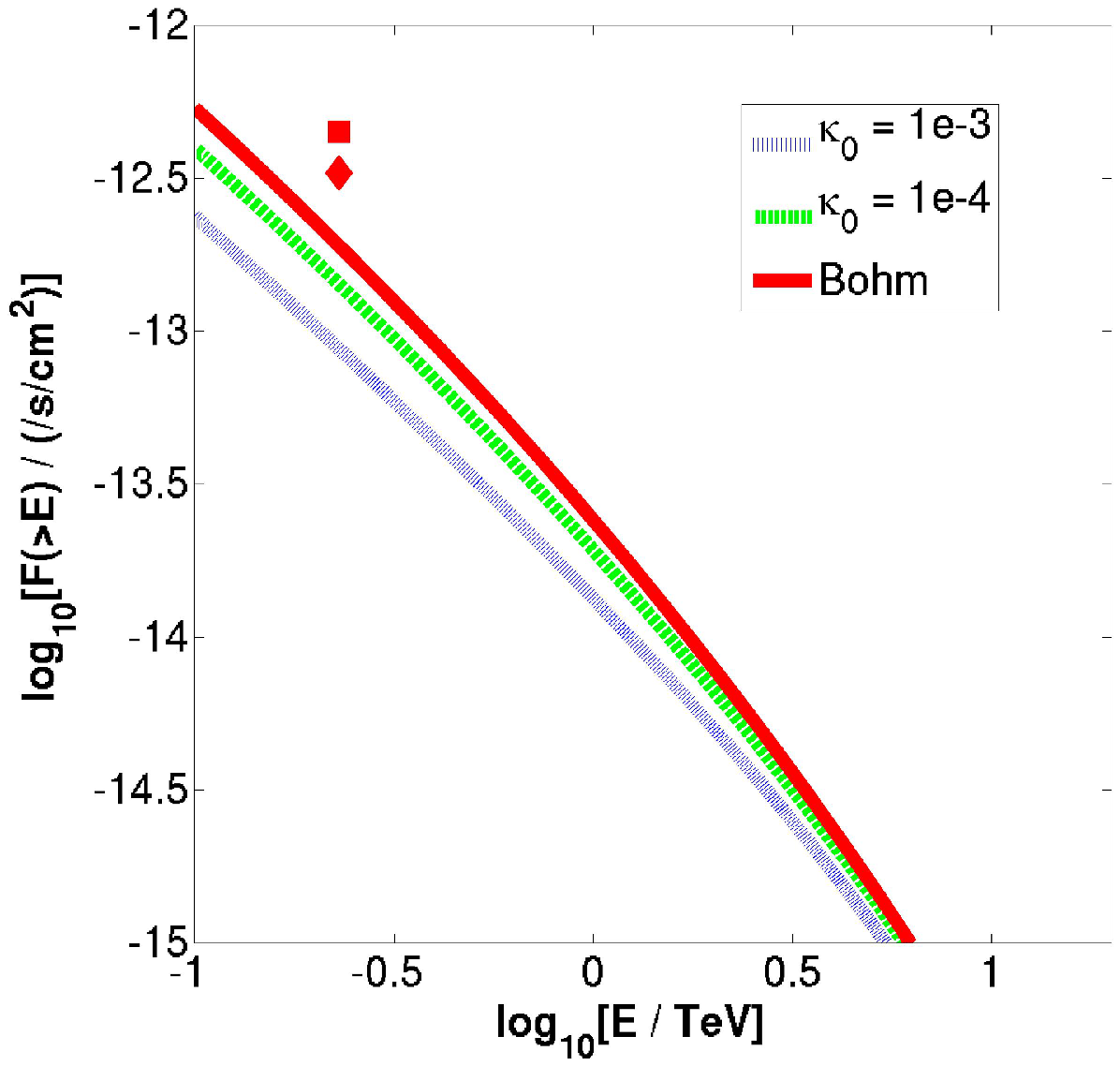}
\caption{\label{fig4} Same as figure~\ref{fig3}, but for the integrated flux. The red square and diamond indicate the extended and point-like H.E.S.S.\ upper limits.}
\end{minipage} 
\end{figure}

We next tested the effect of changing the diffusion coefficient. Figure~\ref{fig3} indicates the average spectra assuming Bohm diffusion (solid red line), a coefficient $\kappa = \kappa_0(E_{\rm e}/1~{\rm TeV})^{0.6}$ with $\kappa_0 =10^{-4}$~kpc$^2$Myr$^{-1}$ (dashed green line), and $\kappa_0 =10^{-3}$~kpc$^2$Myr$^{-1}$ (dotted blue line). The change in spectral shape indicates the different energy dependencies ($\kappa\propto E_{\rm e}^{1.0}$ for Bohm diffusion versus $\kappa\propto E_{\rm e}^{0.6}$). The corresponding integral fluxes are shown in figure~\ref{fig4}. 
%We also indicate a representative error band on the $\kappa_0=10^{-3}$~kpc$^2$Myr$^{-1}$ case of a factor 5. 

%, which would mean that the observational upper limit may yet be satisfied.  
Our next task is a rigorous assessment of the error on the average integral spectrum, taking into account the range of values that each free parameter may assume, for each individual cluster. Preliminary investigations have shown that we can reduce the number of free parameters to five. Varying these parameters on a grid and calculating the average flux value and spread give quite large errors \cite{Venter_HESA}. This process is very time-consuming, given the enormous number of combinations in which a cumulative (or average) spectrum may be obtained (15 GCs, with at least 5 free parameters each).

\section{Conclusions}
\label{sec:concl}
We have described the application of our leptonic GC model to a population of clusters that have been observed, but not detected, in VHE $\gamma$-rays. The fact that upper limits are available presented us with the opportunity to test our model using a sizable sample of clusters. The upper limit for the stacked (average) flux is stringent, and a simple scaling model violated this limit. In contrast, none of the our single-cluster spectral predictions violated the respective upper limits. Our average model spectrum also satisfied the stacked upper limits for the given choice of parameters (table~\ref{tab1}), which we regarded as reasonable. There is, however, considerable uncertainty in the single-GC parameters, so that the intrinsic error on the predicted average spectrum will be quite large. As an example, we showed that different assumptions for the diffusion coefficient lead to significant changes in spectral and flux. One should therefore attempt to reduce the number of free model parameters and also constrain their values so as to decrease the predicted flux error. This will allow one to more robustly test the viability of the MSP scenario for explaining the (non-)detection of TeV flux from GCs.

\ack
This research is based on work supported by the South African National Research Foundation.

%Jain S C, Willander M, Narayan J and van Overstraeten R 2000 {\it J. Appl. Phys}. {\bf 87} 965-78
%\item Selberherr S 1984 {\it Analysis and Simulation of Semiconductor Devices} (Berlin: Springer)


\section*{References}
\begin{thebibliography}{99}
% \bibitem{Abdo09_Tuc} Abdo A A {\it et al.} 2009 {\it Science} {\bf 325} 845--8
\bibitem{Abdo10} Abdo A A {\it et al.} 2010 {\it Astron. Astrophys.} {\bf 524} A75
\bibitem{Abramowski11} Abramowski A {\it et al.} 2011 {\it Astron. Astrophys.} {\bf 531} L18--22
\bibitem{Abramowski13} Abramowski A {\it et al.} 2013 {\it Astron. Astrophys.} {\bf 551} A26
% \bibitem{Abramowski11_NGC6388} Abramowski A {\it et al.} 2011 {\it Astrophys. J.} {\bf 735} 12--9
% \bibitem{Ackermann11_PWN} Ackermann M {\it et al.} 2011 {\it Astrophys. J.}  {\bf 726} 35--51
% \bibitem{Aharonian08_GRB} Aharonian F {\it et al.} 2008 {\it Astron. \& Astrophys.} {\bf 477} 353--63
\bibitem{Aharonian09_Tuc} Aharonian F {\it et al.} 2009 {\it Astron. Astrophys} {\bf 499} 273--7
\bibitem{Alpar82} Alpar M A, Cheng A F, Ruderman M A and Shaham J 1982 {\it Nature} {\bf 300} 728--30
% \bibitem{Anderhub09} Anderhub H {\it et al.} 2009 {\it Astrophys. J.} {\bf 702} 266--9
% \bibitem{Atoyan06} Atoyan A, Buckley J and Krawczynski H 2006 {\it Astrophys. J.} {\bf 642} L153--6
\bibitem{BS07} Bednarek W and Sitarek J 2007 {\it Mon. Not. Royal Astron. Soc.} {\bf 377} 920--30
% \bibitem{Bednarek11} Bednarek W 2011 {\it High-Energy Emission from Pulsars and their Systems: Proc.\ 1st Session Sant Cugat Forum of Astrophys.} ed N Rea and D F Torres pp 185--205 arXiv:1009.1694
\bibitem{Bednarek12} Bednarek W 2012 {\it J. Phys. G: Nucl. Part. Phys.} {\bf 39} 065001
% \bibitem{Buesching08} B{\"u}sching I, Venter C and de Jager O C 2008 {\it Adv. Space Res.} {\bf 42} 497--503
\bibitem{Cheng10} Cheng K S {\it et al.} 2010 {\it ApJ} {\bf 723} 1219--30
\bibitem{Clapson11} Clapson A-C, Domainko W F, Jamrozy M, Dyrda M and Eger P 2011 {\it Astron. Astrophys.} {\bf 532} A47
\bibitem{Domainko11} Domainko W F 2011 {\it Astron. Astrophys.} {\bf 533} L5--8
\bibitem{Eger10} Eger P, Domainko W F and Clapson A-C 2010 {\it Astron. Astrophys.} {\bf 513} A66
\bibitem{Eger12} Eger P and Domainko W F 2012 {\it Astron. Astrophys.} {\bf 540} A17
% \bibitem{Freire11} Freire P C C {\it et al.} 2011 {\it Science} {\bf 334} 1107--10 
% \bibitem{Gehrels05} Gehrels N {\it et al.} 2005 {\it Nature} {\bf 437} 851--4
% \bibitem{Grindlay05} Grindlay J E, Heinke C, Edmonds P D and Murray S S 2001 {\it Science} {\bf 292} 2290--5
% \bibitem{HMZ02} Harding A K, Muslimov A G and Zhang B 2002 {\it Astrophys. J.} {\bf 576} 366--75
\bibitem{HUM05} Harding A K, Usov V V and Muslimov A G 2005 {\it Astrophys. J.} {\bf 622} 531--43
\bibitem{Harris96} Harris W E 1996 {\it Astron. J.} {\bf 112} 1487--8
% \bibitem{Hartwick82} Hartwick F D A, Grindlay J E and Cowley A P 1982 {\it Astrophys. J.} {\bf 254} L11--3
% \bibitem{Hui10} Hui C Y, Cheng K S and Taam R E 2010 {\it Astrophys. J.} {\bf 714} 1149--54
% \bibitem{Kong10} Kong A K H, Hui C Y and Cheng K S 2010 {\it Astrophys. J.} {\bf 71} L36--9
\bibitem{Kopp13} Kopp A, Venter C, B\"usching I and de Jager O C 2013 {\it Astrophys. J.} {\bf 779} 126--37
\bibitem{Lang93} Lang K R 1993 {\it Astrophysical Data: Planets and Stars} (Heidelberg: Springer-Verlag) 257--77  
% \bibitem{KG95} Krockenberger M and Grindlay J E 1995 {\it Astrophys. J.} {\bf 451} 200--9
% \bibitem{McCutcheon09} McCutcheon M {\it et al.} 2009 {\it Proc. 31st ICRC, Lodz, Poland} arXiv:0907.4974
\bibitem{2FGL} Nolan P L {\it et al.} 2012 {\it ApJS} {\bf 199} 31--76
% \bibitem{Ok07} Okada Y, Kokubun M, Yuasa T and Makishima K 2007 {\it Publ. Astron. Soc. Japan} {\bf 59} 727--42
% \bibitem{Pellizoni09} Pellizzoni A {\it et al.} 2009 {\it Astrophys. J.} {\bf 695} L115--9
\bibitem{Pooley03} Pooley D {\it et al.} 2003 {\it Astrophys. J.} {\bf 591} L131--4
% \bibitem{Prinsloo} Prinsloo L P, Venter C, Buesching I and Kopp A {\it these proceedings}
% \bibitem{Ransom08} Ransom S M 2008 {\it 40 Years of Pulsars: Millisecond Pulsars, Magnetars and More, AIP Conf. Ser., ed C Bassa, Z Wang, A Cumming, V M Kaspi} {\bf 983} pp 415--23
% \bibitem{Tam11} Tam P H T {\it et al.} 2011 {\it Astrophys. J.} {\bf 729} 90--7
% \bibitem{Trager95} Trager S C, King I R and Djorgovski S 1995 {\it Astron. J.} {\bf 109} 218--41
% \bibitem{Venter08_conf} Venter C and de Jager O C 2008 {\it AIP Conf. Ser.} {\bf 1085} 277--80
\bibitem{Venter08} Venter C and de Jager O C 2008 {\it Astrophys. J.} {\bf 680} L125--8
\bibitem{Venter09_GC} Venter C {\it et al.} 2009 {\it Astrophys. J.} {\bf 696} L52--5
\bibitem{Venter_HESA} Venter C and Kopp A 2015 {\it Mem.\ S.A.It.}\ {\bf 86} 69--76
% \bibitem{Venter09} Venter C {\it et al.} 2009 {\it Astrophys. J.} {\bf 707} 800--22
% \bibitem{Venter11_Fermi} Venter C, de Jager O C, Kopp A and B{\"u}sching I 2011 {\it 2011 Fermi Symp. Proc. eConf C110509} arXiv:1111.1289
% \bibitem{Verbunt87} Verbunt F and Hut P 1987 {\it IAU Symposium ed D J Helfand and J-H Huang} {\bf 125} 187--96
\bibitem{Wu14} Wu E M H {\it et al.} 2014 {\it Astrophys. J.} {\bf 788} L40--4
\bibitem{Zajczyk13} Zajczyk A, Bednarek W and Rudak B 2013 {\it MNRAS} {\bf 432} 3462--73
\end{thebibliography}
\end{document}